\documentclass[]{spie}  

\usepackage{amsmath,amsfonts,amssymb}
\usepackage{graphicx}
\usepackage[colorlinks=true, allcolors=blue]{hyperref}

\title{High-contrast imaging at first-light of the GMT: the wavefront sensing and control architecture of GMagAO-X}

\author[a,b]{Sebastiaan Y. Haffert}
\author[b]{Jared R. Males}
\author[b]{Laird M. Close}
\author[c]{Maggie Y. Kautz}
\author[b]{Olivier Durney}
\author[b,c,d,e]{Olivier Guyon}
\affil[a]{Leiden Observatory, Leiden University, PO Box 9513, 2300 RA Leiden, The Netherlands}
\affil[b]{University of Arizona, Steward Observatory, Tucson, Arizona, United States}
\affil[c]{Wyant College of Optical Science, University of Arizona, 1630 E University Blvd, Tucson, AZ 85719, USA}
\affil[d]{Astrobiology Center, National Institutes of Natural Sciences, 2-21-1 Osawa, Mitaka, Tokyo, JAPAN}
\affil[e]{National Astronomical Observatory of Japan, Subaru Telescope, National Institutes of Natural Sciences, Hilo, HI 96720, USA}

\authorinfo{Further author information: (Send correspondence to S.Y.H)\\A.A.A.: E-mail: haffert@strw.leidenuniv.nl}

\begin{document} 
\maketitle

\begin{abstract}
The Giant Magellan Adaptive Optics eXtreme (GMagAO-X) instrument is a first-light high-contrast imaging instrument for the Giant Magellan Telescope (GMT). GMagAO-X’s broad wavelength range and the large 25-meter aperture of the GMT creates new challenges: control of all 21.000 actuators; phasing GMT’s segmented primary mirror to nm levels; active control of atmospheric dispersion to sub milli-arcsecond residuals; no chromatic pupil shear to minimize chromatic compensation errors; integrated focal plane wavefront sensing and control (WFS\&C). GMagAO-X will have simultaneous visible and infra-red WFS channels to control the 21.000 actuator DM. The infra-red arm will be flexible by incorporating switchable sensors such as the pyramid or Zernike WFS. One innovation that we developed for GMagAO-X is the Holographic Dispersed Fringe Sensor that measures differential piston. We have also developed several integrated coronagraphic wavefront sensors to control non-common path aberrations exactly where we need to sense them. We will discuss the key components of the WFS\&C strategies for GMagAO-X that address the challenges posed by the first high-contrast imaging system on the ELTs.
\end{abstract}


\section{INTRODUCTION}
\label{sec:intro}
Direct imaging of Earth-like exoplanets is one of the key science goals of the upcoming ground-based telescopes like the Giant Magellan Telescope (GMT) and the European Extremely Large Telescope (ELT) \cite{kasper2021pcs, males2022conceptual}. However, direct imaging of exoplanets is a challenging endeavor. Ground-based telescopes observe astronomical objects through Earth's atmosphere which severely degrades their performance. Adaptive optics (AO) is used to compensate for the atmospheric turbulence \cite{guyon2018exao}. Another major challenge is the large contrast between the exoplanet and its host star. Earth-like planets are easily tens of billions times fainter at separation of a couple times the diffraction limit of the telescopes. Imaging such exoplanets will require extreme high performance adaptive optics systems.

The Giant Magellan Adaptive Optics eXtreme (GMagAO-X) instrument is a high-contrast imaging instrument that is currently being designed for the 24.5-meter GMT \cite{males2022conceptual}. It is designed with the goal of imaging Earth-like planets around nearby stars. The University of Arizona Space Institute (UASI) funded a conceptual design study for GMagAO-X beginning in February 2021. A positive Conceptual Design Review (CoDR) on September 14th, 2021, determined that the project could move into a UASI-funded preliminary design phase. This phase has now been completed with a positive preliminary design review (PDR) in February 2024. The project will now move onto it final design phase. GMagAO-X is currently the only direct imaging instrument under development for any of the large ground-based telescopes that can observe exoplanets in the visible spectral range. Current estimates of its performance show that it could image and characterize a large population (50 to 300) radial-velocity detected planets in reflected light. GMagAO-X will use a 21000 actuator deformable mirror (DM) to achieve high sensitivity at visible wavelengths. The 21000-actuator DM will use novel pupil slicing and recombining techniques to implement a parallel DM where each GMT segment will get its own 3K Boston Micromachine DM\cite{close2022gmagaoxDM}. 

A major challenge will be the wavefront sensing and control architecture for such a large actuator DM. Current cameras put strong limitations on the number of available pixels and read-out speeds. A strong driver in the design of GMagAO-X is to use hardware components that are available today to make sure we can build the instrument whenever the instrument gets greenlighted. This proceeding will describe an approach that is being investigated for GMagAO-X that works with current hardware. In Section 2 we will describe the overall architecture. Each of the sub-components of the full loop are described in Sections 3, 4 and 5.

\section{REQUIREMENTS}
The top-level science requirement of GMagAO-X is to be able to detect reflected light planets. This is a challenging requirement and it is driving almost all of the lower-level requirements. The flowdown from the science requirement result in the following wavefront sensing and control requirements,

\begin{itemize}
    \item The wavefront sensor shall have a limiting magnitude of at least 12 in I-band (GMAGX-SCI-007).
    \item The AO system shall deliver a Strehl of 70\% on a 5th magnitude star at H-alpha (GMAGX-007).
    \item The wavefront sensor shall have a framerate of at least 2 kHz (GMAGX-4).
    \item The wavefront sensor shall reconstruct at least all spatial frequencies up to 182 cycles across the pupil (GMAGX-96). This is set by the 7x 3K DMs of GMagAO-X.
    \item The wavefront sensor shall measure differential piston to an accuracy of 1 nm rms for 8th magnitude or brighter (GMAGX-97). 
    \item The wavefront sensor system shall be able to measure differential atmospheric dispersion (GMAGX-98).
\end{itemize}

\section{Wavefront sensing architecture}
The initial WFS\&C architecture for GMagAO-X used a triple stage system where an initial low-order K-band wavefront sensor would close the loop on the GMagAO-X woofer \cite{haffert2022visible}. Then a high-sensitivity wavefront sensor, such as the Zernike wavefront sensor (ZWFS), would take over and control the high-order DM. All the modeled end-to-end simulations used novel non-linear reconstructors to increase the dynamic range \cite{haffert2024into}. However, the end-to-end closed-loop simulations using I-band for the high-order loop did not always converge with a ZWFS. The simulations did converge at J and H-band. However, those led to phase wrapped solutions. The need for phase unwrapping will reduce the magnitude limit because those generally require high SNR. From these results we determined that the dynamic range of the Zernike wavefront sensor was too small to guarantee the required performance. The multi-stage approach did not meet our requirements. 

\subsection{Choice of wavefront sensor}

\begin{figure}
    \begin{center}
        \includegraphics[width=\textwidth]{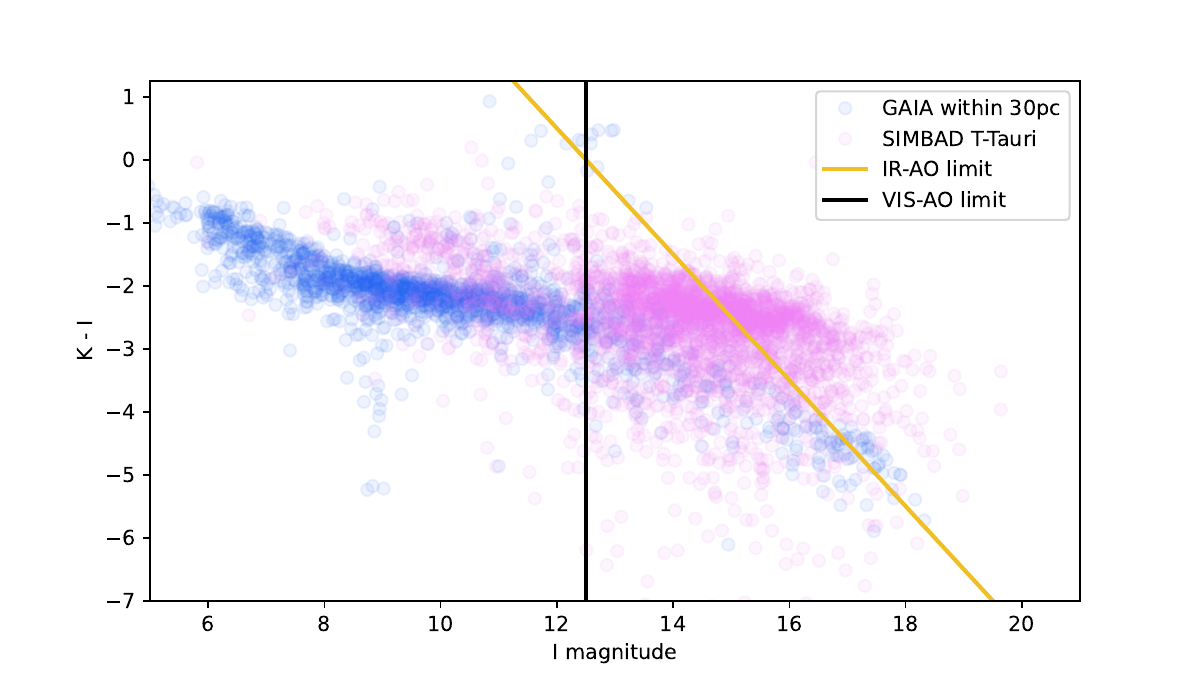}
	\end{center}
    \caption{The color-color diagram of potential targets for GMagAO-X. These targets have been selected based on their visibility from Las Campanas Observatory. Many targets are too faint for visible wavefront sensor. However, their brightness at redder wavelengths makes them suitable for infrared wavefront sensing.}
    \label{fig:visir_targets} 
\end{figure}

One of the risk reduction principles of GMagAO-X is to try to solve as many problems as possible with currently available technology. This lead to the choice to use the (modulated) pyrmaid wavefront sensor. This is wavefront sensor has a lot of heritage within the GMagAO-X design team due to the experience with MagAO and MagAO-X. Finally, the sensitivity gain of the ZWFS over the (un)modulated PWFS is not that big due to the aperture shape of GMT. The unmodulated PWFS has a sensitivity of 1.4 and the ZWFS a sensitivity of 1.5 to 1.7. This marginal gain in sensitivity requires a lot of additional complexity in the wavefront reconstruction and trades in robustness. We do want to keep the option for newer Fourier WFS architectures that might outperform the pyramid in the future \cite{landman2022joint, haffert2023reaching}. Therefore, the wavefront sensor design will be a flexible Fourier Filtering WFS with a stage that can change the focal plane optics. The sensitivity of the modulated PWFS is lower than the unmodulated PWFS. However, we are currently developing a deep-learning based wavefront reconstruction algorithm to enable wavefront control with the unmodulated PWFS \cite{landman2024making}. The internal source experiments have been extremely successful and now we are moving towards on-sky experiments to demonstrate the performance.

The conventional PWFS has a prism with four faces that create four pupils. For GMagAO-X, the base line wavefront sensor will be the (modulated) three-sided pyramid wavefront sensor (3PWFS). The 3PWFS is more robust against read noise than the four-sided PWFS (4PWFS). And, it is also easier to manufacture a three-sided pyramid than a four-sided pyramid. This should result in a smaller ridge on the tip. The 3PWFS performs equally to the 4PWFS in the bright star case \cite{schatz2021three, chambouleyron2023modeling} and it slightly pushes the guide star magnitude to lower magnitudes. There is already ample experience in running a 3PWFS \cite{schatz2022three}.

\subsection{Field of view requirement}
The high-order parallel DM of GMagAO-X can create spatial frequencies up to about 182 cycles / pupil. The wavefront sensor must be oversampled to ensure accurate reconstruction and reduced aliasing errors. Super-resolution techniques could be used to increase the number of reconstructed modes. However, such techniques suffer in performance loss compared to oversampled wavefront sensors. The extreme AO requirements of GMagAO-X push us towards minimal wavefront error for brighter targets. Controlling all these modes also require a precise registration of the DM with respect to the wavefront sensor. All of these problems are reduced when oversampling the wavefront sensor. The GMagAO-X will use a 20\% oversampling which is similar to that of MagAO-X. This results in 220 pixels across the pupil.

GMagAO-X uses a knife-edge mirror to separate the incoming and outgoing beams from the parallel DM. This results in a rectangular field of view. The wavefront sensor drives the minimal size because a all spatial frequencies up to 220 cycles/pupil must be passed through. The field of view requirement is,
\begin{equation}
    \mathrm{FoV} = \pm 110 \lambda_{\mathrm{max}} / D.
\end{equation}
Here $\lambda_{\mathrm{max}}$ is the largest wavelength in the bandpass of GMagAO-X and $D$ the telescope diameter. The largest wavelength of GMagAO-X is 1.9 um and the telescope diameter is 24.5m. This results in a minimal field of view of 3.5 arcseconds.

\subsection{Multi-channel wavefront sensors}
The wavefront sensor band determines what targets are available for AO on GMagAO-X. Adding IR WFS capabilities would add a substantial number of nearby stars for reflected light science. Additionally, it would almost triple the amount of T-Tauri stars we can observe. This is important for the other major science case of accreting proto-planets. Adding IR capabilities would more than tripple the number of targets. Figure \ref{fig:visir_targets} shows an overview of potential targets. GMagAO-X shall have both visible and IR wavefront sensing capabilities. No high-speed camera  with sub-electron read noise currently has the capability to sense light from 600 nm to 1900 nm (the spectral bandwidth of GMagAO-X). Therefore, GMagAO-X will have two wavefront sensor channels; a visible channel covering 600 nm to 900 nm and an IR channel that will cover 900 to 1900 nm. Each channel will use a dedicated double pyramid prism where each pyramid is made of a different type of glass. This double pyramid approach was used to create more achromatic response \cite{tozzi2008double}. We optimized the type of glass and the pyramid apex angles of the visible pyramid and the IR pyramid to create an achromatic response over their respective bandpasses. The current designs of the pyramid prisms can be found in Table 1 and Table 2.

\begin{table}[]
\label{tab:irpwfs}
\caption{The design parameters of the IR PWFS of GMagAO-X.}
\begin{tabular}{l|l|l|l}
Parameter & Value & Unit & Comment  \\ \hline \hline
Wavelength range & 900 to 1900 & nm & IR spectral bandpass  \\
Angle 1          & 30 & deg & fixed angle  \\
Material 1       & S-BAH11 &  &   \\
Angle 2          & 29.8 & deg & optimized angle  \\
Material 2 & S-LAL58 &  &   \\
Pupil separation  & 1.557 &  pupil units &   \\
Input focal ratio  & 60 &  & based on GMagAO-X optical design \\
Pixels across pupil  & 220 & pixels & set by sampling requirements \\
Pixel size  & 24 & um & set by SAPHIRA camera  \\
Final pupil size & 5.28 & mm &   
\end{tabular}
\end{table}

\begin{table}[]
\label{tab:vispwfs}
\caption{The design parameters of the VIS PWFS of GMagAO-X.}
\begin{tabular}{l|l|l|l}
Parameter & Value & Unit & Comment  \\ \hline \hline
Wavelength range & 500 to 900 & nm & VIS spectral bandpass  \\
Angle 1          & 30 & deg & fixed angle  \\
Material 1       & S-NSL3 &  &   \\
Angle 2          & 29.718 & deg & optimized angle  \\
Material 2 & K10 &  &   \\
Pupil separation  & 1.557 &  pupil units &   \\
Input focal ratio  & 60 &  & based on GMagAO-X optical design \\
Pixels across pupil  & 220 & pixels & set by sampling requirements \\
Pixel size  & 6.5 & um & set by TELEDYNE Kinetix  \\
Final pupil size & 1.43 & mm &   
\end{tabular}
\end{table}

\subsection{Differential piston sensing}
Differential piston is a complicated mode to sense for the pyramid wavefront sensor especially when the differential piston errors are bigger than $\pm \lambda/2$ \cite{hedglen2022lab}. Therefore, we are using the Holographic Dispersed Fringe Sensor (HDFS) for phasing the segments of the GMT \cite{haffert2022phasing}. The HDFS uses a pupil plane hologram to interferometrically combine pairs of segments onto separate locations in the focal plane. This allows us to measure the differential phase between all pairs of segments. The dispersion that is intrinsic in these holograms allows us to disentangle large differential piston signals. A prototype of the HDFS has recently been tested on-sky and it successfully demonstrated phasing to the white light fringe (Kautz et al. in prep). The response of this sensor will also be used to measure residual differential atmospheric dispersion. The details can be found in \cite{twitchell2024ADC}.

\section{SEMI-ANALYTICAL MODEL PERFORMANCE}
\begin{figure}
    \begin{center}
        \includegraphics[width=0.7\textwidth]{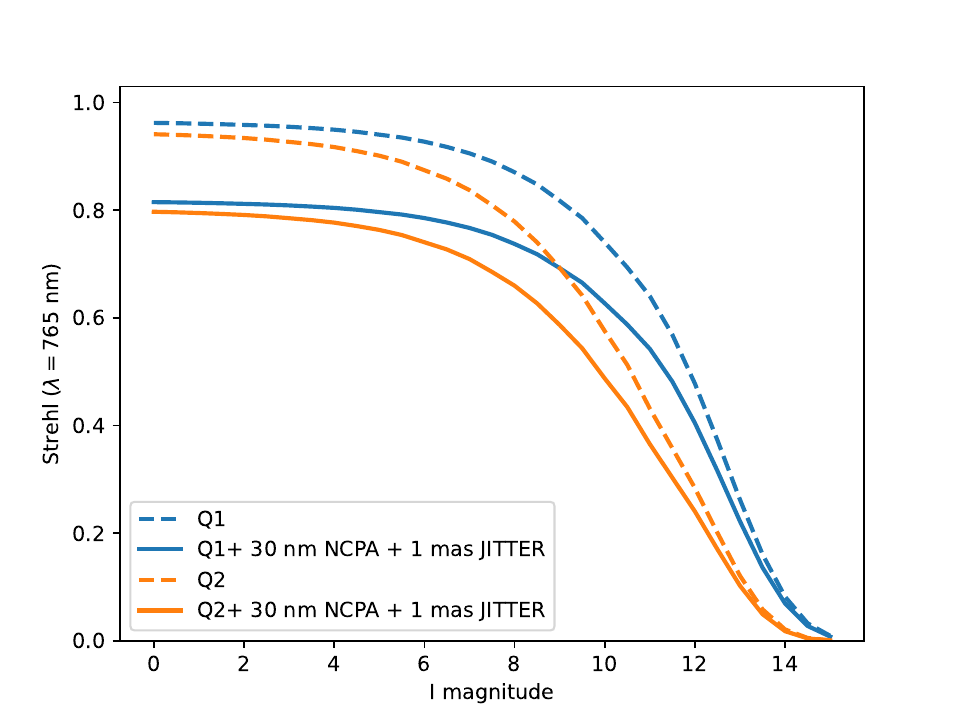}
	\end{center}
    \caption{The GMagAO-X Strehl as a function of stellar magnitude in I band. The Strehl is shown for various conditions. The dashed lines show the Strehl when including the fitting error, photon and detector noise and the aniso-server error. The solid lines show a more realistic Strehl when also including 30 nm of uncorrectable NCPA and 1 mas of residual jitter. The blue curve shows the performance for the top 25 percentile conditions and the orange curve is for the top 50 percentile conditions (median conditions).}
    \label{fig:strehl} 
\end{figure}
Full end-to-end simulations of an high-order AO system like GMagAO-X are very expensive in terms of computational time and complexity. Therefore, we used a semi-analytical approach to estimate the instruments performance \cite{jolissaint2010synthetic}. The underlying assumption of all temporal evolution is Taylor's frozen flow. Using this approach it becomes possible to replace any temporal evolution by a spatial shift. The time evolution of any parameter can then be replaced by a shifted version, $\varphi(\vec{r}, t)=\varphi(\vec{r} - \vec{v}t)$. This time evolution underpins the semi-analytical model and allows us to express an adaptive optics system as a series of spatial filters that act on the turbulence Power Spectral Density (PSD). The AO system is controlled using integral control, $\mathrm{DM}_{i+1} = \mathrm{DM}_{i} - g \varepsilon_i$. Here, $\varepsilon_i$ is the wavefront error at time step $i$ and $\mathrm{DM}_i$ is the DM shape at time step $i$. The feedback gain $g$ is a function of spatial frequency and was optimized for each spatial mode, stellar magnitude and loop frequency \cite{gendron1994astronomical, jolissaint2010synthetic, males2021mysterious}.

\section{Focal plane wavefront sensing and control}
Control of non-common path aberrations is crucial for high-contrast imaging. MagAO-X has pioneered the architecture where the coronagraph has a dedicated DM that is invisible to the high-order wavefront sensor. This allows us to decouple NCPA offloading and atmospheric turbulence control. This approach where we separate the control loops was successfully tested on MagAO-X (Haffert et al. in prep, Kueny et al. this proceedings). Based on this succes, the GMagAO-X coronagraph arm shall have a 3K DM in or near the pre-apodizer pupil plane. This Non-common path corrector DM (NCPCDM) shall correct the NCPA and use dark-hole digging algorithms like iEFC \cite{haffert2023implicit}.

There will be multiple options for focal plane wavefront control for GMagAO-X. These are based on where the light is split off and the type of coronagraph that is used.

\subsection{Parametric Phase Diversity}
Parametric phase diversity uses a parametric optical model together with several defocused images to measure the phase and amplitude aberration. This algorithm is currently running on MagAO-X on the internal source. The focus offsets are created with the NCPCDM. Parametric phase diversity has been extremely sucesful on MagAO-X. The instrumental NCPA were cleaned up to the nm level within the control radius of the DM \cite{van2021characterizing}. The algorithm has also been demonstrated on-sky with MagAO-X where it was used to clean up NCPA on stellar targets. Parametric phase diversity uses the science camera itself without any coronagraph.

\begin{figure}
    \begin{center}
        \includegraphics[width=0.9\textwidth]{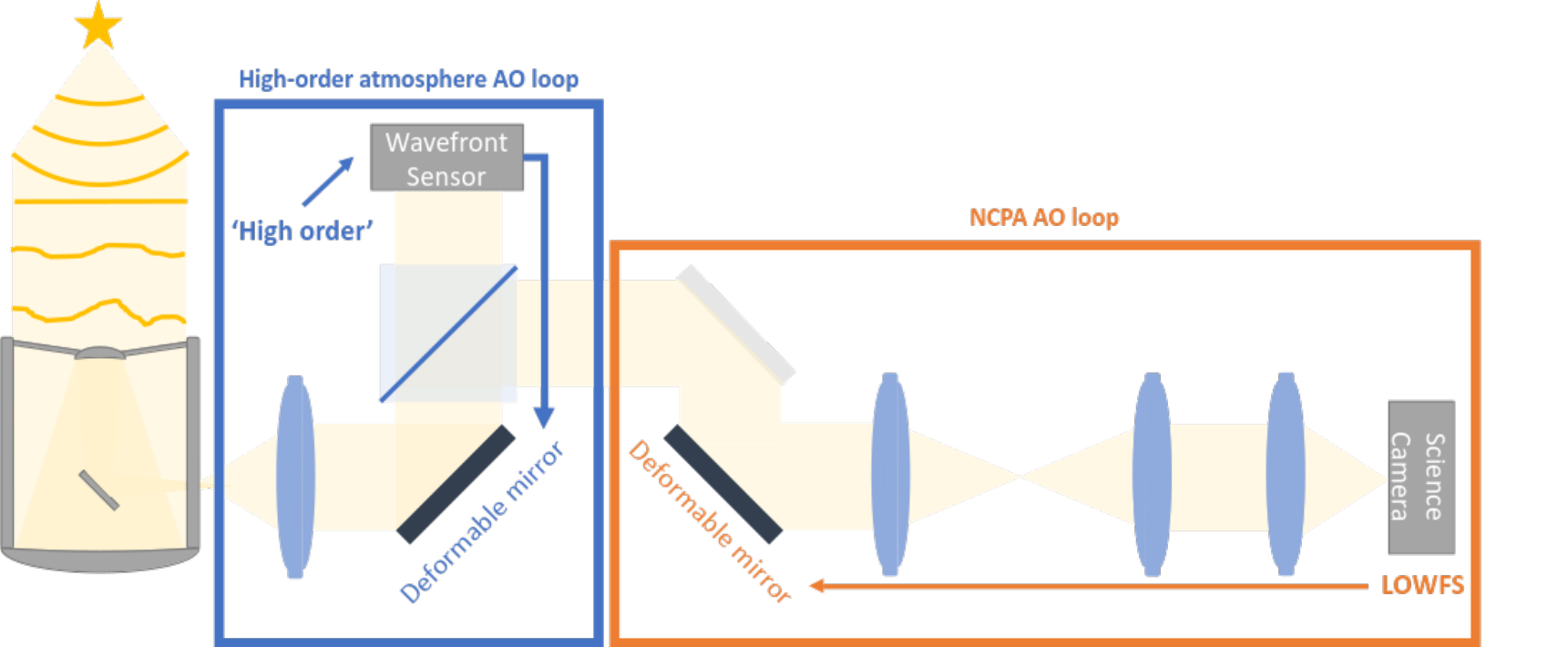}
	\end{center}
    \caption{The schematic layout for parametric phase diversity.}
    \label{fig:fdpr} 
\end{figure}

\subsection{Coronagraphic low-order WFS}
The first option for coronagraphic wavefront sensing uses the light that is reflected of the focal plane mask of the coronagraph. This type of wavefront sensing is enabled by making all focal plane mask reflective instead of opaque. So any part that needs to remove light reflects instead of absorbs. The reflected part of the PSF is reimaged onto a separate focal plane. A high-speed camera with low-read noise will then use the PSF for LDFC-like wavefront control \cite{miller2017spatial} or linearized phase diversity. This approach is called Coronagraphic Low Order WFS (CLOWFS).

\begin{figure}
    \begin{center}
        \includegraphics[width=0.9\textwidth]{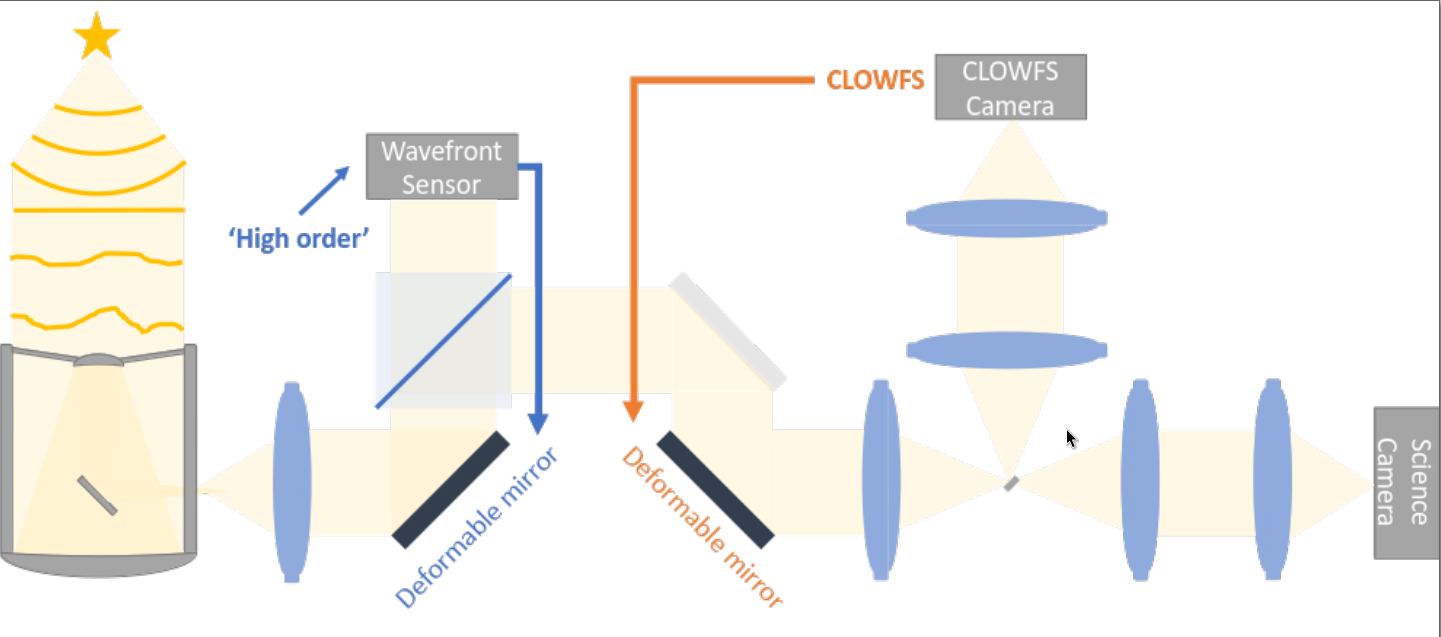}
	\end{center}
    \caption{The schematic layout for coronagraphic low-order wavefront sensing.}
    \label{fig:clowfs} 
\end{figure}

\subsection{Lyot low-order WFS}
The second option for coronagraphic wavefront sensing uses the light that is reflected of the Lyot stop of the coronagraph. This type of wavefront sensing is enabled by making all Lyot stop reflective instead of opaque. So any part that needs to remove light reflects instead of absorbs. This is necessary for coronagraphs that use phase-shifting focal plane masks because these don't have any part can reflect light for the CLOWFS. The reflected light outside the Lyot stop is imaged onto a dedicated high-speed camera. The low-order wavefront errors can then be measured using Lyot Low-order WFS (LLOWFS) \cite{singh2014lyot}.

\begin{figure}
    \begin{center}
        \includegraphics[width=0.9\textwidth]{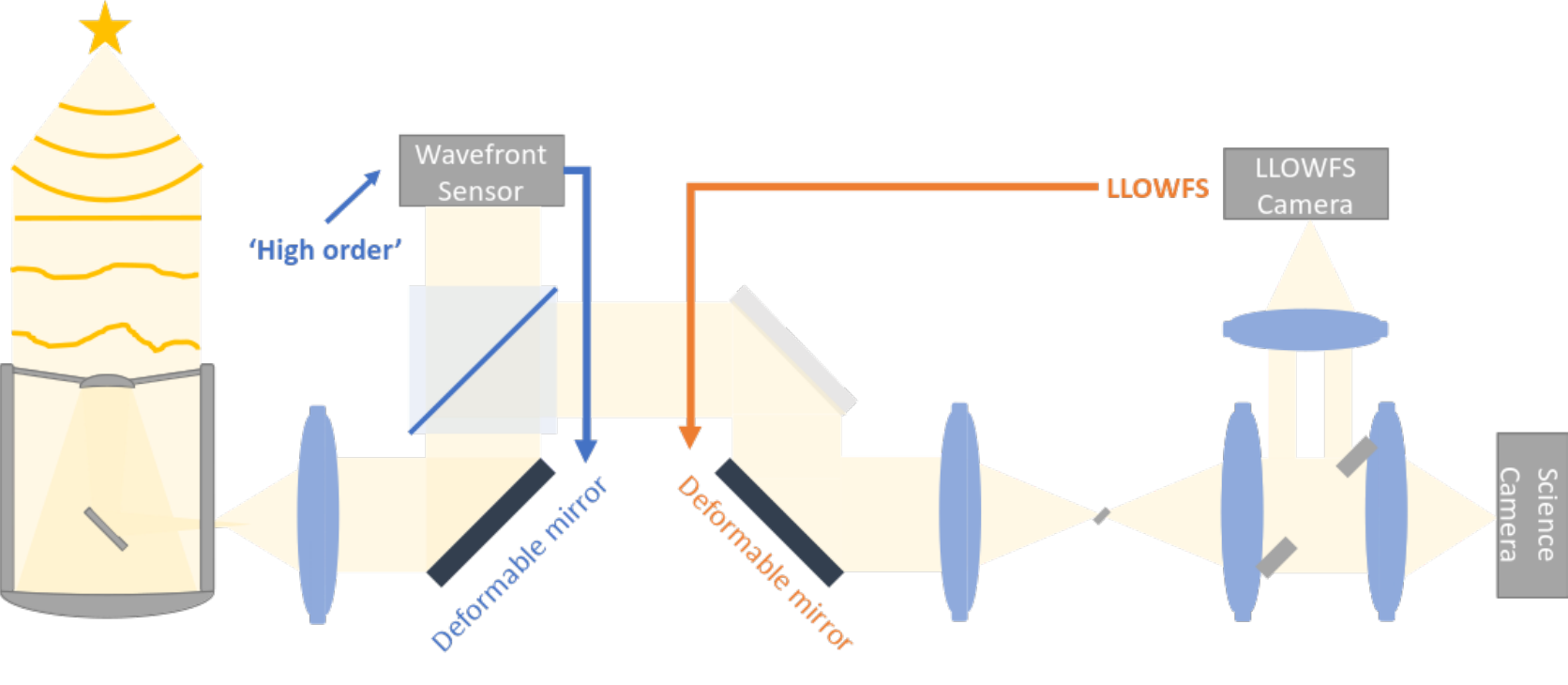}
	\end{center}
    \caption{The schematic layout for Lyot low-order wavefront sensing.}
    \label{fig:llowfs} 
\end{figure}

\subsection{Speckle Nulling}
The final set of focal plane wavefront control algorithms are the speckle nulling algorithms. These use direct feedback from the science focal planes to measure the electric field using either pair-wise probing \cite{haffert2023implicit} or the Self-Coherent Camera \cite{haffert2023integrated}. The focal plane measurements are then used to directly null the electric field and create dark hole regions in the focal plane.

\section{CONCLUSION}
The design of GMagAO-X is now well under way. We are solving new issues as this is the first Extreme AO system that is being developed for the new generation of extremely large telescopes. The proposed design of GMagAO-X is able to meet the required specifications and is using technology which is currently available. The current design is set to meet the design specifications of 70\% Strehl at H$\alpha$ on a 5th magnitude star.

\acknowledgments 
Support for this work was provided by the generous support of the Heising-Simons Foundation and the University of Arizona Space Institute.

\bibliography{report} 
\bibliographystyle{spiebib} 

\end{document}